\documentclass[12pt]{article}

\usepackage{url}
\usepackage{graphicx}
\usepackage{xcolor}
\usepackage[numbers]{natbib}
\usepackage{amsfonts}
\usepackage{hyperref}
\usepackage{caption}
\usepackage{authblk}

\usepackage[left=1.91cm,right=1.91cm,bottom=2.54cm,top=2.54cm]{geometry}

\def\tblhead#1{\hline\\[-9pt]#1\\\hline\\[-9.75pt]}
\def\lastline{\hline}
\providecommand{\keywords}[1]{\textbf{\textit{Keywords:}} #1}
\date{}

\begin{document}

\title{Sample size determination via learning-type curves}

\author[$\dag$1]{Alimu Dayimu\thanks{Corresponding author Alimu Dayimu ad938@medschl.cam.ac.uk}}
\author[2]{Nikola Simidjievski\thanks{These authors contributed equally to this work}}
\author[3,4]{Nikolaos Demiris}
\author[4]{Jean Abraham}
\affil[1]{Cambridge Clinical Trials Unit, Department of Oncology, University of Cambridge, Cambridge, UK}
\affil[2]{Department of Computer Science and Technology, University of Cambridge, Cambridge, UK}
\affil[3]{Department of Statistics, Athens University of Economics and Business, Athens, Greece}
\affil[4]{Cambridge Clinical Trials Unit, Cambridge University Hospitals NHS Foundation Trust, Cambridge, UK}
\affil[5]{Cambridge Breast Unit, Cambridge University Hospitals NHS Foundation Trust, Addenbrooke's Hospital, Cambridge, UK}

\maketitle

\begin{abstract}
{This paper is concerned with sample size determination methodology for prediction models. We propose to combine the individual calculations via learning-type curves. We suggest two distinct ways of doing so, a deterministic skeleton of a learning curve and a Gaussian process centred upon its deterministic counterpart. We employ several learning algorithms for modelling the primary endpoint and distinct measures for trial efficacy. We find that the performance may vary with the sample size, but borrowing information across sample size universally improves the performance of such calculations. The Gaussian process-based learning curve appears more robust and statistically efficient, while computational efficiency is comparable. We suggest that anchoring against historical evidence when extrapolating sample sizes should be adopted when such data are available. The methods are illustrated on binary and survival endpoints.
}
\end{abstract}

\section{Introduction}
\label{sec1}

Risk prediction models are routinely used in healthcare and medical research \citep{qrisk2,altman2007prognostic} to inform the diagnosis and/or prognosis of clinical events~\citep{steyerber2019, hemingway2013prognosis}. Constructing risk prediction models relies on different modelling approaches, ranging from well-established statistical methods to more recent machine learning algorithms. However, irrespective of the underlying modelling methodology, leveraging data with an appropriate sample size for developing such models is imperative for achieving robust and accurate predictive performance on a given task, such as predicting binary, continuous or time-to-event outcomes. Therefore, accurate sample size calculations are necessary for the development phase, facilitating reliable and accurate prediction models.

Approaches to sample size calculation can differ based on the predictive task and modelling strategy as well as the underlying study design. In this paper, we focus on risk prediction models that address binary and time-to-event outcomes, but the techniques we develop are also directly applicable to continuous outcomes. This problem is distinct from sample size calculation, where one is interested in estimating the accuracy of a diagnostic test since that target accuracy may be independent of the sample size. In contrast, when developing prediction models, performance may vary with sample size, possibly due to including covariates that may improve predicting performance as the sample size increases, or because a more elaborate model may be employed. 

A typical approach to determining an adequate sample size is factoring the number of predictor variables, such as ensuring at least ten events per predictor~\citep{moons2014critical}. Whilst simple, such criteria do not consider the predictors' type, magnitude and possible values (e.g. categorical variables may require more events) -- often leading to poorly fitted models that fail to generalise well to out-of-sample data~\citep{vansmeden2019, van2016no}. In response, recent simulation studies~\citep{vansmeden2019} point to additional and necessary requirements to inform the sample size estimation, which relates to the choice of the modelling strategy and their expected out-of-sample performance. Riley et al \citep{riley2019minimum} worked on models where $R^2$-type measures are applicable and do incorporate the expected model performance, the number of (candidate) predictors and the outcome prevalence in the target population into the sample size calculation. The performance of a prediction model varies depending on the modelling strategy and variables included in the model, such as relationships between predictors and response variable(s).

In a clinical research setting, the statistician is typically interested in (1) calculating the optimal sample size associated with the study design before the data collection stage OR (2) assessing the feasibility of an ongoing study after a certain number of samples has been accrued. To address these issues, our study focuses on estimating the predictive performance of a modelling strategy at the design stage, possibly utilising external data if applicable. Specifically, we are concerned with (i) selecting a strategy and the associated sample size for developing a prediction model, (ii) assessing the feasibility of an ongoing study after a certain number of samples has been accrued. The technique used to develop a risk prediction model may differ depending on the method used and the variables included. For example, prediction effects and collinearity issues may affect the downstream performance of the resulting model. To assess the operating characteristics of a strategy we propose a learning curve approach that provides performance estimates at different sample sizes and can leverage historical evidence.

A learning curve may be based on an inverse power law model that describes the performance of a modelling strategy as a function of the sample size \citep{mukherjee2003,viering2021shape} used for developing a prediction model. Related work in different scenarios \citep{Figueroa2012,Beleites2013, christodoulou2021adaptive} estimate the expected performance of a linear classifier for different sample sizes. However, prediction models can be highly unstable, particularly when conditioned upon small sample sizes used at the modelling stage. Therefore one may benefit by leveraging information from publicly available data from related studies in order to inform the operating characteristics of a strategy's performance for substantially larger sample sizes. In particular, during the study planning stage, this may help determine the optimal sample size needed to achieve a desired accuracy in a robust manner.

In this work, we study a learning curve-based framework to estimate the expected predictive performance of a modelling strategy. We propose and evaluate approaches that utilise external data from similar studies, leading to robust estimates for a given sample size. Specifically, we employ a bootstrap strategy on external data to derive performance estimates at different (incremental) sample sizes. We treat these estimates as data and estimate the parameters of a learning-curve which in turn estimates the expected performance of the underlying model. We employ four statistical models and evaluate the proposed approaches on a series of real-world experiments, predicting binary and time-to-event outcomes. We show that extrapolation may be unstable in scenarios with limited available data and illustrate that incorporating external data via evidence synthesis can be highly beneficial at the design stage. The paper is structured as follows. Section 2 discusses the developed learning-curve framework that estimates the prediction accuracy of prediction models at a given sample size. In Section 3, we apply this method to publicly available data with binary and survival outcomes and Section 4 concludes with a discussion.

\section{Methods}\label{sec2}

We introduce a learning curve meta-modelling framework to estimate the expected performance of risk prediction models (and associated strategies) conditional on sample size. Specifically, we fit a learning curve using estimates of predictive performance derived from different modelling strategies over repeated random draws of incremental sample sizes. To achieve more stable performance, we leverage information from external data and anchor against those for extrapolating to larger sample sizes. The main elements of our proposed framework are presented below.

\subsection{Learning curves}\label{learncurve}

Learning curves model the relationship between the predictive performance (PP), $Y_{pp}$, of a prediction model as a function of the sample size $n$ used for constructing it \citep{cortes1993learning}. Here we employ a \emph{regular} learning curve whose expected predictive performance is non-decreasing with sample size: $\mathbb{E} \left[ Y_{pp}(n+1) \right] \geq \mathbb{E} \left[ Y_{pp}(n) \right]$. While there are several curve types that can be utilised for modelling such behaviour, in this paper we focus on the power law model since it typically provides very good fit \citep{guModelling,brumen2014} while retaining a natural interpretation of its parameters. The functional form is given by:

\begin{equation} \label{eqln}
Y_{pp}(n)=f(n;a,b,c)=(1-a)-b n^{-c}
\end{equation}
where the parameter $a$ denotes the minimum achievable error (ranging from 0 to 1), $b$ denotes the learning rate and $c$ the decay rate with $c \in (0,1)$. For the predictive tasks considered in this paper $Y_{pp}$ is bounded in $[0,1]$ and the maximum is achieved at $(1-a)$ for $n \to \infty$. 

In this work, we build on two distinct but related approaches for estimating a learning curve: (i) a frequentist approach fitted via Nonlinear Least Squares (NLS), and (ii) a Bayesian approach where the learning curve is modelled as a Gaussian Processes (GP). The NLS is implemented using the Levenberg-Marquardt algorithm \citep{more1978levenberg} for weighted NLS to fit the learning curve given in equation (\ref{eqln}). The $a$ and $c$ parameters are $logit$ transformed for stability. 

The Bayesian approach is based upon a Gaussian processes assuming the model performance is distributed as:

\begin{equation} \label{eqgp}
Y_{pp}(n) \sim \mathcal{N}(g(n), \sigma_y^2)
\end{equation}
where $\sigma_y$ denotes the standard deviation and $g(n)$ is a Gaussian process $$g(n) \sim \mathcal{ GP }(\mu(n),\Sigma)$$ with mean $\mu(n) = (1-a)-b n^{-c}$, and covariance matrix $\Sigma$. We consider $\Sigma_{ij} = \phi^2\exp{( -\rho |n_i - n_j|^2)}$ and complete the model with equation (\ref{eqgp}). We place beta priors on $a$ and $c$ and a normal distribution on $b$. Note that, since $Y_{pp} \in [0, 1]$ we could have transformed $Y_{pp}$ to $\mathbb{R}$, but we found this to be unnecessary in our application.

While superficially there are similarities between our work and the methods of \citep{Figueroa2012} and \citep{Beleites2013} who also use NLS to construct a deterministic learning curve, our approach facilitates several additional methodological features mostly related to the Gaussian process and anchoring against external data for both linear and non-linear models. 

\subsubsection*{Other learning curve components}

The estimated learning curve parameters generally depend on the data, the sampling method, the predictive tasks, and the modelling approach.
An essential component of developing risk prediction models is the choice of appropriate evaluation criteria that assess the predictive and/or discriminative ability of the developed models for clinical decision support/making. Commonly used metrics include $R^2$, the Brier score and the Area Under the Receiver Operating Characteristics Curve (AUC) among others. In this study, we measure the predictive performance $Y_{pp}$ using the C-statistic, commonly used in clinical studies, to evaluate the accuracy of risk prediction models \citep{tripod}. For binary outcomes, it is a proxy to the AUC \citep{bamber1975area}. For time-to-event outcomes we use the censoring-adjusted C-statistic \citep{unoC2011} as a prediction measure. 

A crucial aspect for the model's generalisability and downstream predictive performance is the underlying modelling methodology. Risk prediction models typically build on well-established approaches from statistical modelling and machine learning such as: (group) lasso \citep{lasso,grouplasso}, support vector machines \citep{vapnik} as well as ensemble methods \citep{breimanRF, friedmanGBM, van2007super,wolpert1992stacked}, which achieve high predictive performance by combining multiple predictors. The choice of an appropriate modelling methodology can rely on many different factors that relate to the study design, dataset properties and the predictive task at hand and the methods we propose may be used for any such method.

\subsection{Algorithm}

Fitting a learning curve of a model's predictive performance, $Y_{pp}$, as a function of the sample size requires a series of performance estimates derived under a series of $m$ sub-samples with sizes of $\{s_1,s_2,\dots,s_m\}$, sampled from the total sample $N$ with size $s_m$ and this sub-sample referred as $n_m$. The sub-samples $n_m$ are sampled randomly and vary in size, increasing from samples with sizes as small as $s_1=50$ to a maximum of samples with size $s_m=N$. 

In this work, we sample $m=50$ such samples, which we then use for training and evaluating predictive models and, in turn, fitting the learning curve using the $m=50$ performance estimates. Note that, for each $m$ we repeatedly sampled $k$ times without replacement. Specifically, for each $n_{mk}$ subsample we estimate a set of model performance estimates $Y_{{pp}_{mk}}$, where $n_m$ denotes a particular sample with size $s_m$ from the set $m$, and $k$ denotes the repeated random draws performed, $k \in \{1, 2, \dots, 100\}$. We develop and validate the modelling strategy via a stratified split of each $n_{mk}$ into $70\%$ training and $30\%$ validation data, the latter being used to obtain $Y_{{pp}_{mk}}$. More formally, the procedure is as follows:

\fbox{
    \parbox{\textwidth}{
\begin{enumerate}
    \item Select the number of sample sizes $m$ and fit a learning curve.
    \item For each size $s_m$ of the set $m$:
    \begin{enumerate} 
    \item Randomly sample $n_{mk}$ of size $s_m$ from the complete data of size $N$ without replacements.
    \item Perform a randomly stratified 70/30 split of $n_{mk}$ into $D_{train}$ for training and $D_{test}$ for validating the modelling strategy, respectively. This was repeated for 100 times.
    \item Fit a model with a pre-specified modelling strategy using the training set $D_{train}$. 
    \item Validate the fitted model in the validation sample $D_{test}$ and calculate the predictive performance $Y_{{pp}_{mk}}$ on $D_{test}$.
    \item Repeat the previous steps for $k=100$ times for every size $s_m$ 
    \end{enumerate} 
   	\item Fit a learning curve to using the obtained $Y_{{pp}_{mk}}$ estimates.
\end{enumerate}
}}
\subsection{Transfer learning strategy}

Extrapolating beyond the observed range of the data, typically in scenarios with limited sample size, can lead to highly unstable predictions and should be done with caution. In such cases anchoring upon data from an external study allows for robust extrapolation to large sample sizes. In particular at the study planning stage one may fit a learning curve, $f(n)_{ext}$, to external but related data. The degree of relatedness may be assessed using commonly used methods. These can vary from trivial similarity such as the same (say binary) outcome, to disease type to more stringent similarity based on inclusion-exclusion criteria.


In the case of estimating the predictive performance of a modelling strategy via NLS, we assume the learning curves fitted on the target and external data have a similar shape but a different scale. Therefore, we first pre-fit a learning curve on external data $f(n)_{ext}$ and on the small-sample target data $f(n)_{target}$. For the estimation, we transfer information on the learning rate $b$ and the decay rate $c$ from the external data, while learning the increment on $a$ from the target data. The estimated predictive performance can be calculated by following for a given sample size $n'$
$$\mathbb{E}(Y_{pp})=(1-a_{target})-b_{ext} n'^{-c_{ext}}$$
where $a_{target}$ is estimated from the target data, while $b_{ext}$ and $c_{ext}$ estimated from the external data. 

The Gaussian processes naturally facilitate the synthesis of different sources of evidence. We use the posterior of $b$, $c$ and $\phi$ estimated from external data as prior when fitting the learning curve to the target data. The predicted performance of $Y_{pp}$ at a sample $n_i$ is used to predict the accuracy of the chosen modelling strategy at a larger sample size for the target data. Given the marginal estimates of $\mu_p$ at $n_p$ and $\mu_o$ at $n_o$, the predicted value $y_p$ is given by:
$$
y_p|y_o\sim \mathcal{N}(\mu_p+\Sigma_{po}\Sigma_{oo}^{-1}(y_o-\mu_o),\Sigma_{pp}-\Sigma_{po}\Sigma_{oo}^{-1}\Sigma_{op})
$$
where $y_o$ denotes the observed performance and $\Sigma$ the covariance matrix. The subscript $o$ denotes the observed data and $p$ denotes the data points to be predicted. In contrast to NLS, using the GP means that the uncertainty around the predicted value can naturally be obtained via the covariance matrix.

\section{Experimental Design}
\label{sec3}

Our application focuses on two clinical settings: (i) estimating the sample size needed at the study design stage and (ii) evaluating the feasibility of achieving a specific performance for an ongoing study with limited samples accrued. We demonstrate our framework using two distinct breast cancer datasets as described below.

\subsection{Breast cancer use cases}

Consider a typical scenario in clinical trials where an investigator aims to design a study to explore different modelling strategies' predictive abilities. In this context, we focus on predicting breast cancer outcomes and select the Molecular Taxonomy of Breast Cancer International Consortium (METABRIC) \citep{metabric2012a} study as external data. METABRIC is an extensive dataset, consisting of detailed genomic profiles and clinical data from breast-cancer patients. In particular, the genomic data includes mRNA expression and Copy number alterations (CNA), while the clinical data encompasses patien
t outcomes and treatment responses from patients across various different subtypes.

For our limited (target) data, we select a subset of the Memorial Sloan Kettering Cancer Center (MSK) breast-cancer data \citep{msk2018}. The complete dataset consists of genomics profiles (CNA) obtained from targeted sequencing of tumour- and normal-sample pairs from 1,918 Breast cancers from 1,756 patients. The genomic data is paired with clinical data, that includes patient outcomes and treatment responses. In our analysis, we experiment with a subset of the MSK dataset, by attempting to extend it to a larger sample sizes by anchoring it to the METABRIC data. The clinical and genomics data for both METABRIC and MSK datasets are publicly available and were obtained from cBioPortal \citep{cerami2012cbio, gao2013integrative}.

We consider both binary (five-year survival status) and time-to-event (overall survival) outcomes. The models incorporate variables reported by Margolin et al. \cite{margolin2013systematic}. In the case of METABRIC this includes age at diagnosis, tumour size, number of positive lymph nodes, tumour grade, oestrogen receptor (ER) status, progesterone receptor (PR) status, human epidermal growth factor receptor 2 (HER2) status, radiotherapy, and hormone therapy. After excluding missing observations, the METABRIC dataset includes 1978 participants, with 427 (21.6\%) death events within the first five years and 1144 (57.8\%) over the complete follow-up period, resulting in a median survival of 13 years. In terms of analysis, we followed Curtis et al.~\citep{metabric2012a} and used a pre-selected set of the input CNA features. Specifically, we used the most significant \textit{cis}-acting genes that are significantly associated with CNAs determined by a gene-centric ANOVA test. To simplify the computational analysis, we selected the genes with the most significant Bonferroni adjusted p-value from the Illumina database containing $30566$ probes. After missing-data removal, the input data sets consisted of $1000$ CNA categorical features.

In the case of  MSK, clinical variables include age at diagnosis, tumour stage, ER status, HER2 status, and PR status. The MSK dataset, after removing missing observations, comprises 1640 participants; 175 (10.7\%) died within 5 years, and 343 (20.9\%) overall, with a median survival of 14.8 years. We present the Kaplan-Meier curves of both datasets in Figure S1 in the Appendix. We assess accuracy using the C-statistics, AUC for binary outcomes, and Uno's C at 10 years for time-to-event outcomes.

\subsection{Learning curve by sample size}

We used the METABRIC data to explore the baseline behaviour of the learning curve over different total sample sizes $N\in\{100, 150, 300, 450, 600, 900, 1200, 1500, 1978\}$. We used logistic regression and included the clinical variables and a total of $m=50$ data points with $s_1=50$ being the smallest sample size of $n_1$. Both NLS and GP were used to fit a learning curve. To further investigate the impact of number of data points on fitting learning curve, a series of $m\in\{5, 10, 20, 30, 40, 50\}$ were explored at different total sample sizes $N\in\{100, 300, 600\}$ with METABRIC data with the same modelling strategy described above. 

\subsection{Leveraging multi-modal data and models}

In clinical research it is vital to select the modelling strategy, variables and prediction model, that is able to achieve good and robust performance with as small a sample size as possible. Under such circumstances, this performance may be further improved by integrating different data modalities collected from the same patients. We explore the accuracy of different models in predicting the METABRIC 5 year survival status using (1) clinical data only and (2) clinical data plus copy number alteration (CNA) variables. The clinical variables included in the prediction model were outlined in 3.1. In this setting we used all the METABRIC data and a total of $m=50$ data points with the smallest sample size $s_1=50$ of the sample $n_1$.

We evaluate prediction models developed using four different approaches: (1) Elastic net (ElNet)~\citep{ellasticnet} (2) Support vector machines (SVMs)~\citep{vapnik} (3) Random Forests (RF)~\citep{breimanRF} and (4) Gradient Boosted Trees (LGBM)~\citep{friedmanGBM, NIPS2017_LightGBM}. All models, from each of the four methods, have their hyper-parameters optimised and are calibrated. For the hyper-parameter optimisation procedure we use Bayesian optimisation search. Specifically, before performing $k=100$ training repetitions, we perform one extra repetition at each sample size, which we use for cross-validated search over hyper-parameters of a given model for optimal C-statistic. We use the resulting hyper-parameter values for the subsequent repetitions. Each model, at every repeat, is calibrated based on Platt's logistic model~\citep{Platt99} using internal validation. This is largely redundant for logistic regression but we retain the same approach for consistency.

\subsection{Sample size extrapolation}

We consider a scenario where limited data are available and a feasibility analysis is planned: the study may terminate early or recruit more patients if the expected performance of the prediction model seems feasible and appropriate. It is hard to fit a robust model with a small sample size so we may borrow information from external data as described above. We illustrate this task using a subset of the MSK study as target data and information from METABRIC as an anchor for extrapolation. Logistic regression and the Cox model were selected as prediction models using the available clinical variables. We assume an interim analysis to evaluate the feasibility of the proposed modelling strategy will be performed after limited samples become available from the MSK study. We choose $m=20$ data points and $s_1=30$ as the smallest sample size, assuming a total of $N\in\{50, 80, 100, 150, 300, 450\}$ were available from the MSK study. The $s_1=30$ chosen in this example to allow for 20 data points when the total sample size $N=50$

\section{Results}
\label{sec4}
\begin{figure}
  \centerline{\includegraphics[width=1\linewidth]{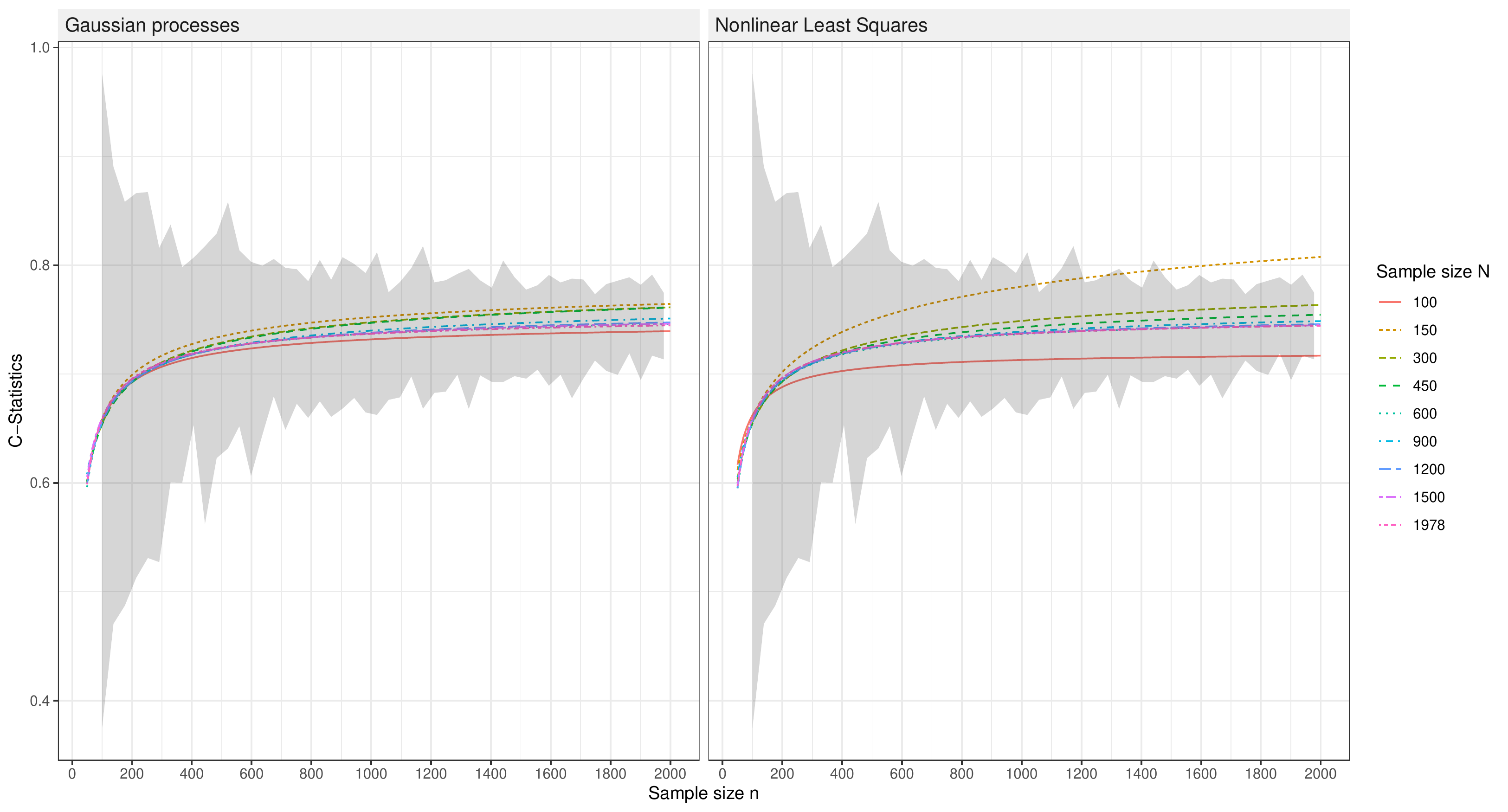}}
  \caption{Predicted C-statistics of learning curves of METABRIC data alone with GP and NLS of logistic regression for binary outcomes at different sample size (lies) and 95\% confidence interval of the raw results using all samples (shaded area)}
  \vspace*{-3pt}
  \label{fig:fig1}
\end{figure}

Figure ~\ref{fig:fig1} presents the two (NLS and GP) C-statistic learning curves as a function of sample size. Both GP and NLS work well with most predicted C-statistics falling within the 95\% CIs across sample sizes. The GP learning curve appears more robust, especially for sample sizes larger than 100, while the NLS method stabilises only after 300 sample size. The shape of both learning curves was similar and the improvement in C-statistics saturated at sample sizes higher than 1000. It is apparent that a C-statistics of 0.8 (or more) is not feasible based on the current modelling strategy and the problem at hand. 


Table ~\ref{tab:table1} presents the point estimates and standard error of the coefficients of both learning curves. The standard errors are higher for the NLS curve and the point estimates stabilise at larger sample sizes suggesting that the GP curve is statistically more efficient. In this settings, a total of $m=50$ data points were used. Table A1 and Figure S2 in the Appendix reveals that the proposed approach is reasonably robust to smaller series of sample sizes (data points) such as $m=5$ for example, with the GP curve being more efficient.

\begin{center}
  \captionof{table}{Point estimation and standard error (SE) of GP and NLS at different total sample size
  \label{tab:table1}}
  \footnotesize
  \begin{tabular}{@{}lccccccc@{}}
  \tblhead{
  &\multicolumn{3}{c}{\textbf{Gaussian Process}} && \multicolumn{3}{c}{\textbf{Nonlinear Least Squares}} \\ \cline{2-4}\cline{6-8}
  \textbf{N} & \textbf{a (SE)}  & \textbf{b (SE)}  & \textbf{c (SE)} && \textbf{a (SE)}  & \textbf{b (SE)}  & \textbf{c (SE)}}
100 & 0.25 (0.013) & 2.07 (0.387) & 0.67 (0.050) && 0.28 (0.724) & 2.58 (11.838) & 0.82 (10.137) \\ 
  150 & 0.21 (0.023) & 1.48 (0.267) & 0.52 (0.068) && 0.00 (850735.729) & 0.82 (0.292) & 0.19 (3.085) \\ 
  300 & 0.20 (0.009) & 1.16 (0.107) & 0.45 (0.031) && 0.19 (0.291) & 1.04 (0.374) & 0.42 (0.588) \\ 
  450 & 0.20 (0.012) & 1.12 (0.150) & 0.44 (0.046) && 0.22 (0.093) & 1.41 (0.428) & 0.53 (0.372) \\ 
  600 & 0.24 (0.004) & 2.00 (0.185) & 0.64 (0.026) && 0.24 (0.035) & 2.22 (0.583) & 0.67 (0.311) \\ 
  900 & 0.22 (0.006) & 1.20 (0.160) & 0.50 (0.038) && 0.23 (0.029) & 1.51 (0.291) & 0.57 (0.204) \\ 
  1200 & 0.23 (0.003) & 1.72 (0.198) & 0.60 (0.031) && 0.24 (0.014) & 2.21 (0.345) & 0.66 (0.166) \\ 
  1500 & 0.24 (0.003) & 1.67 (0.175) & 0.60 (0.028) && 0.24 (0.008) & 2.41 (0.298) & 0.70 (0.133) \\ 
  1978 & 0.24 (0.002) & 2.26 (0.265) & 0.68 (0.028) && 0.24 (0.003) & 2.52 (0.164) & 0.71 (0.068) \\ 
  \lastline
  \end{tabular}
\end{center}

\begin{figure}
  \centerline{\includegraphics[width=1\linewidth]{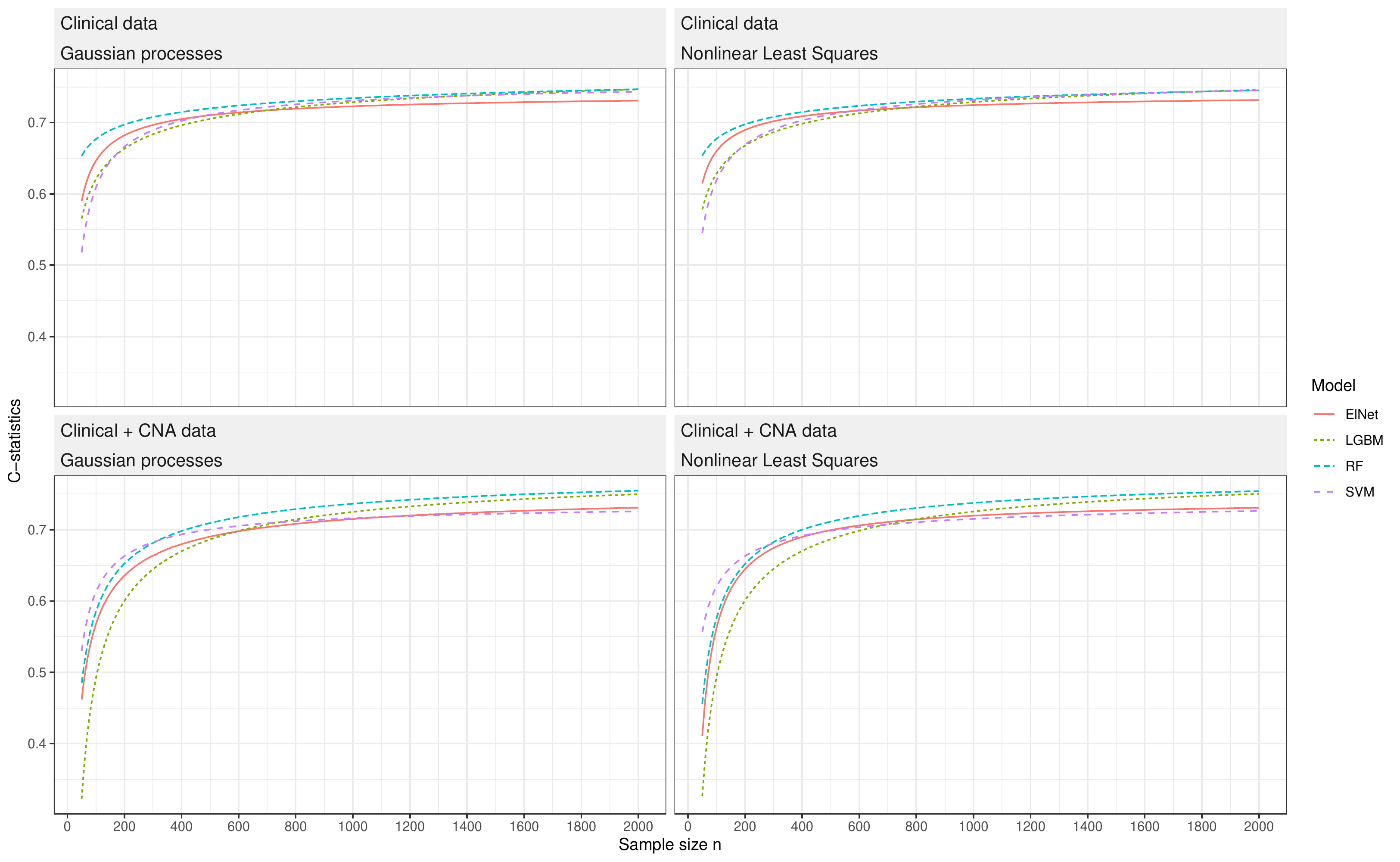}}
  \caption{Predicted C-statistics of learning curves of METABRIC alone with GP and NLS of Elastic Net (ElNet), light gradient-boosting machine (LGBM), random forest (RF), support vector machine (SVM) trained and evaluated in both uni-modal (clinical only data) and multi-modal (clinical + CNA data) scenarios.}
  
  \vspace*{-3pt}
  \label{fig:fig2}
\end{figure}

Figure ~\ref{fig:fig2} presents the learning curves with the predicted C-statistics of different modelling strategies in an uni-modal (clinical data alone) and multi-modal (clinical data plus CNA data) setting. The NLS and GP-based learning curves were similar across sample sizes. Including the CNA data in the prediction model did not considerably improve the accuracy in this use case. All models performed similarly at large sample sizes while the LGBM slightly under-performed at smaller samples. The ElNet and SVM reached a plateau at 1000 samples while accuracy improvement was minimal when more data were added to the other models.

\begin{figure}
  \centerline{\includegraphics[width=1\linewidth]{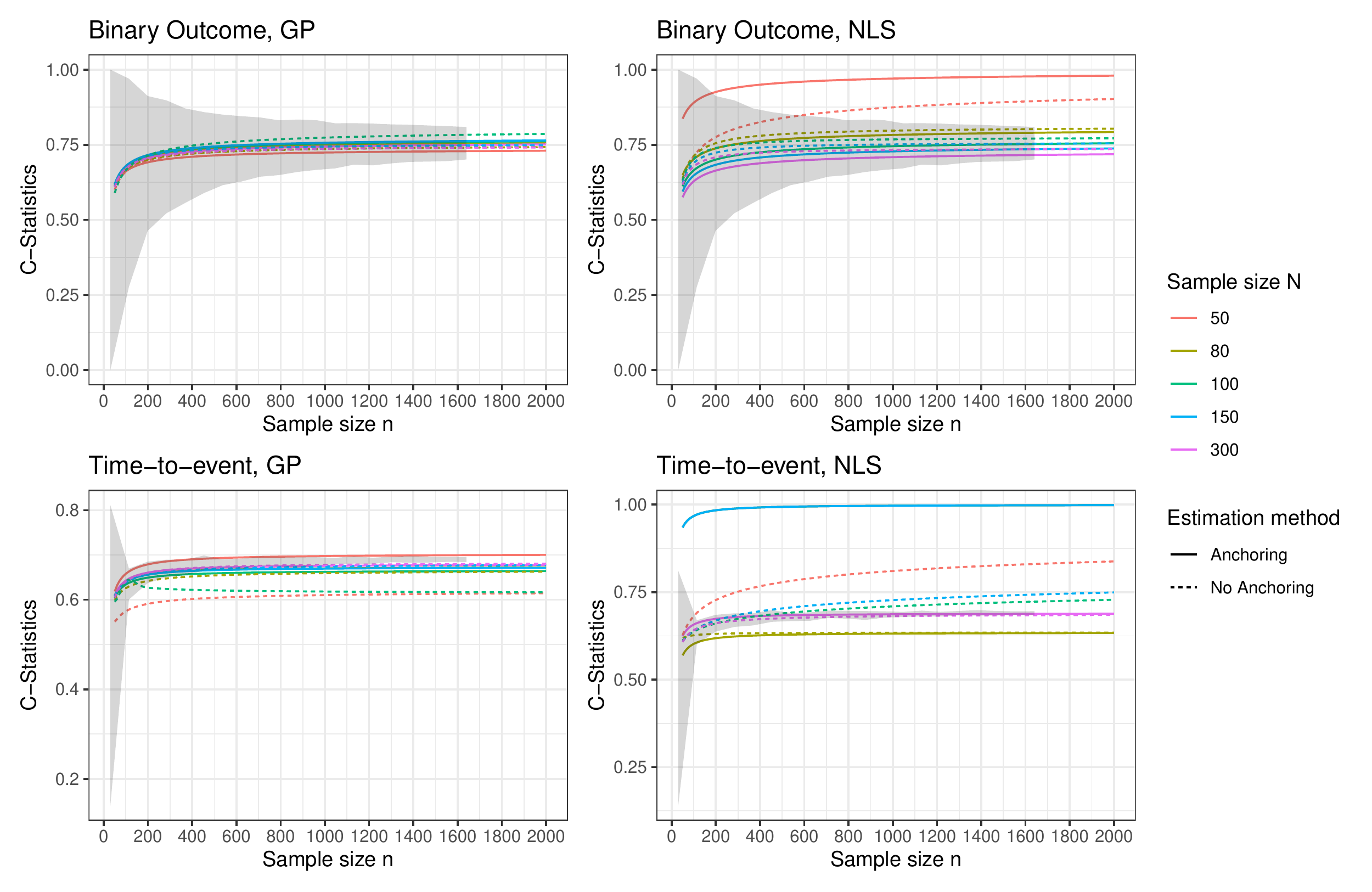}}
  \caption{Predicted C-statistics of learning curve of MSK with GP and NLS with (solid line) and without (dashed line) anchoring METABRIC data for binary and time-to-event outcomes using logistic regression and survival model, respectively. The shaded area is the 95\% confidence interval of C-statistics from repeated random draws using all MSK samples without extrapolation.}
  \vspace*{-3pt}
  \label{fig:fig3}
\end{figure}

Figure ~\ref{fig:fig3} presents the behaviour of NLS and GP learning-curve models on the MSK cohort, with and without anchoring to the METABRIC data. For binary outcomes, both GP and NLS can produce curves that are effective and stable, independently of anchoring. Specifically, while NLS generally stabilises at sample sizes of 80 and larger, GPs exhibit reasonable performance at sample sizes of 50. However, for survival outcomes, anchoring can lead to additional stability, especially at small sample sizes, in the case of GP learning curves. This, however, is not the case for NLS curves, which tend to diverge at sample sizes smaller than 300.

In a case study, the MSK study in this example, an interim analysis was performed after accruing a total of 50 samples. The investigator was interested in predicting the accuracy of the model at 1640 samples (full MSK samples). The logistic model predicted C-statistics of 0.90 (95\% CI 0.56-1.23) and 0.74 (95\% CI 0.69-0.79) for the NLS and GP methods, respectively, without anchoring. When anchoring against METABRIC data, the predicted C-statistics were 0.98 and 0.73 (95\% CI 0.72-0.73) for the NLS and GP methods. The raw C-statistics in the random draws from the total MSK samples were 0.76 (95\% CI 0.70-0.81). For the survival model without anchoring, the predicted C-statistics were 0.83 (95\% CI 0.49-1.16) and 0.61 (95\% CI 0.54-0.68) for the NLS and GP methods, respectively. When anchoring against METABRIC data, the predicted C-statistics were 1.00 and 0.70 (95\% CI 0.69-0.70) for the NLS and GP methods. The raw C-statistics in the random draws from the total MSK samples were 0.69 (95\% CI 0.64-0.73).

\section{Discussion}
\label{sec5}

We proposed a flexible and robust approach to sample size prediction based on learning curves. The goal is to estimate the expected predictive performance of a model (and a modelling strategy) at a given sample size in the design stage. Our approach has several practical benefits which can be employed for model selection via analysing the performance of different prediction models, and, as such, used in an ongoing study to inform their feasibility or futility. Specifically, we propose two variants of a learning curve, showing that the Bayesian GP-based version can achieve better performance at small to moderate sample sizes. The obtained learning curves of the NLS were considerably different at sample size $N=100$ and $N=150$, which is reflected in the NLS parameter estimation. At a small sample size, the C-statistics were increasing before the inflection point. The parameters of the NLS were bounded to the parameter space by parameterisation and may result in a converging issue. These differences, on the other hand, were negligible at large sample sizes. We also showed that the stability and robustness of the developed learning curves could be further improved by anchoring the fitting process against a larger study on related data, where available.  

We explored the value of adding different data modalities such as genetic data, which in our experiments led to small gains in predictive performance. Implicitly and more broadly, our approach allows for assessing the gain and the cost of collecting distinct data modalities at the planning stage. As a result, it appears sensible for one to modify the prioritisation of the research question in order to maximise the expected benefit.  

The proposed approach is general, modular and flexible. We used the C-statistic for assessing the predictive performance, but different scoring rules, like the Brier score, may also be used. Moreover, our approach allows for principled analysis of modelling strategies. As the sample size increases, different statistical and machine learning algorithms may be investigated for model development. Within-model variable selection can also be explored in a similar manner. We employed a 70/30 train/test split when obtaining predictive performance estimates but different rules, such as 80/20 splits, may also be used. 

Finally, in the presence of limited data, we suggest anchoring on a larger study of similar data with a GP approach. The use of external data is routinely employed in different extrapolation settings like survival analysis and makes intuitive sense. It may be hard to select the external data and quantify the similarity to the target data. The choice of external data can rely on expert (e.g. clinical) opinion or by inspecting the definition of the populations, such as inclusion-exclusion criteria and the relatedness of the endpoints. In the absence of high-quality clinical trial data, real-world evidence could be a useful alternative. From the theoretical standpoint one could define an appropriate loss function that assesses fidelity to the distinct data sources via appropriate weights and this is the subject of current research.

\section*{Acknowledgements}

We thank Simon Bond for thoughtful feedback on this work. NS and JA acknowledge financial support from Mark Foundation for Cancer Research and Cancer Research UK Cambridge Centre [C9685/A25177]. JA also acknowledges financial support from Cancer Research UK, the National Institute for Health and Care Research and from Cambridge Biomedical Research Centre.

\newpage
  
\small
\bibliographystyle{unsrt}
\bibliography{ref}

\newpage
\appendix
\makeatletter
\renewcommand \thesection{S\@arabic\c@section}
\renewcommand\thetable{S\@arabic\c@table}
\renewcommand \thefigure{S\@arabic\c@figure}
\setcounter{figure}{0}
\setcounter{table}{0}
\makeatother

\section*{Appendix}

\section{Kaplan Meier curves of METABRIC and MSK data\label{app1}}
\begin{figure}[!h]
  \centerline{\includegraphics[width=1\linewidth]{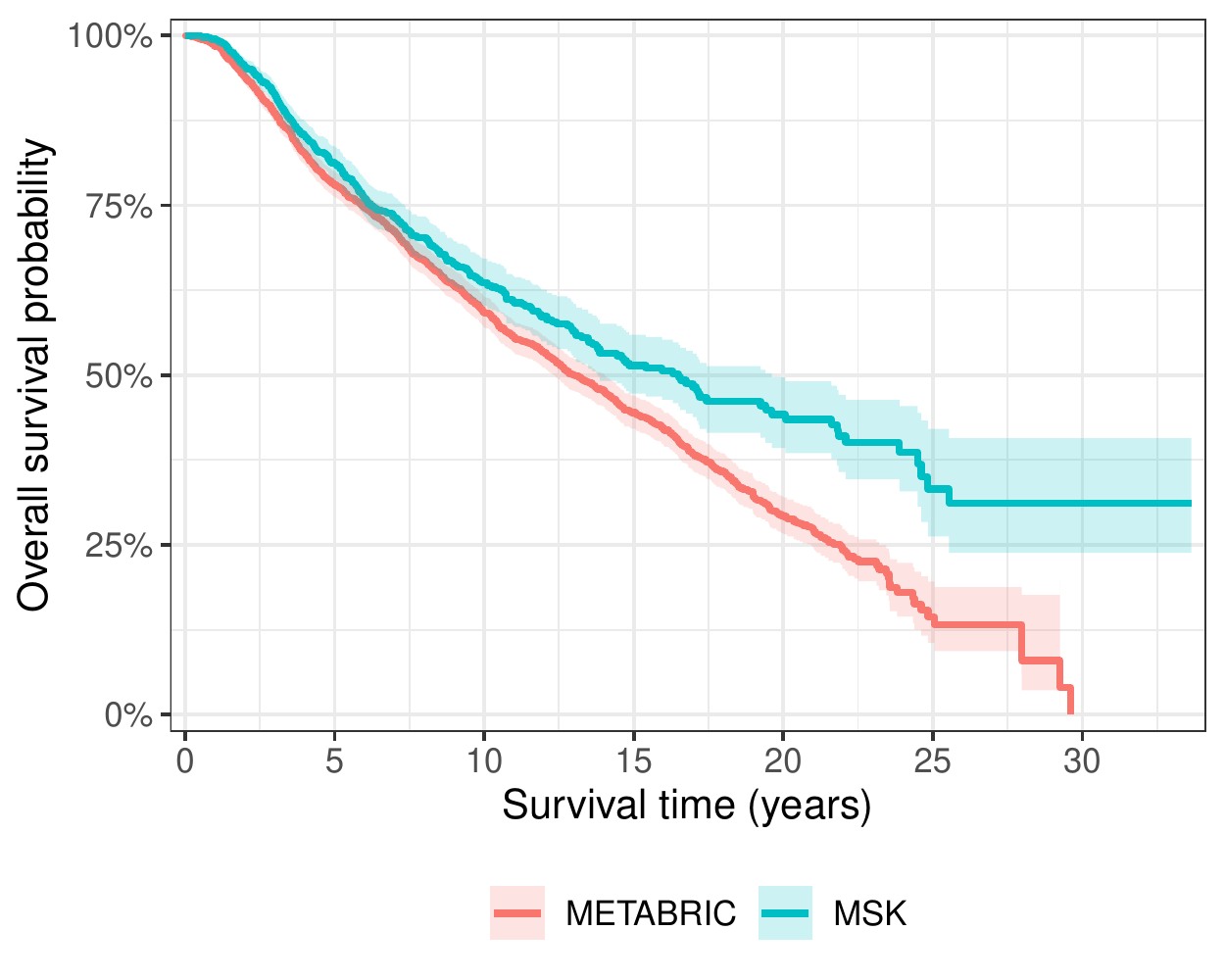}}
  \caption{Kaplan Meier curve of METABRIC and MSK data}
  \label{figfigs1}
\end{figure}

Figure ~\ref{figfigs1} presents the Kaplan Meier curves of METABRIC and MSK data.

\section{Impact of number of data points $m$ used for estimating the learning curves\label{app2}}

\vspace*{12pt}

In our main analysis, we use $m=50$ data points, which denote samples with different sizes, for estimating the learning curves. To further investigate the impact of number of data points on the stability of the fitted learning curves, we perform analyses of different sets $m\in\{5, 10, 20, 30, 40, 50\}$, at a varying total sample size $N\in\{100, 300, 600\}$. We use METABRIC data and evaluate a modelling strategy with a logistic regression model trained and evaluated only with clinical data on a task of predicting a binary outcome.

Figure ~\ref{figfigs2} presents the two (NLS and GP) C-statistic learning curves as a function of sample size. Both GP and NLS work well with most predicted C-statistics falling within the 95\% CIs for most sample sizes. Table ~\ref{tabtabs1} presents the point estimation of learning learning curve model. The GP model appears more robust even with only 5 data points and 100 total sample size. GP and NLS had similar results after the total sample size were 300 or more.  

\begin{center}
  \includegraphics[width=1\linewidth]{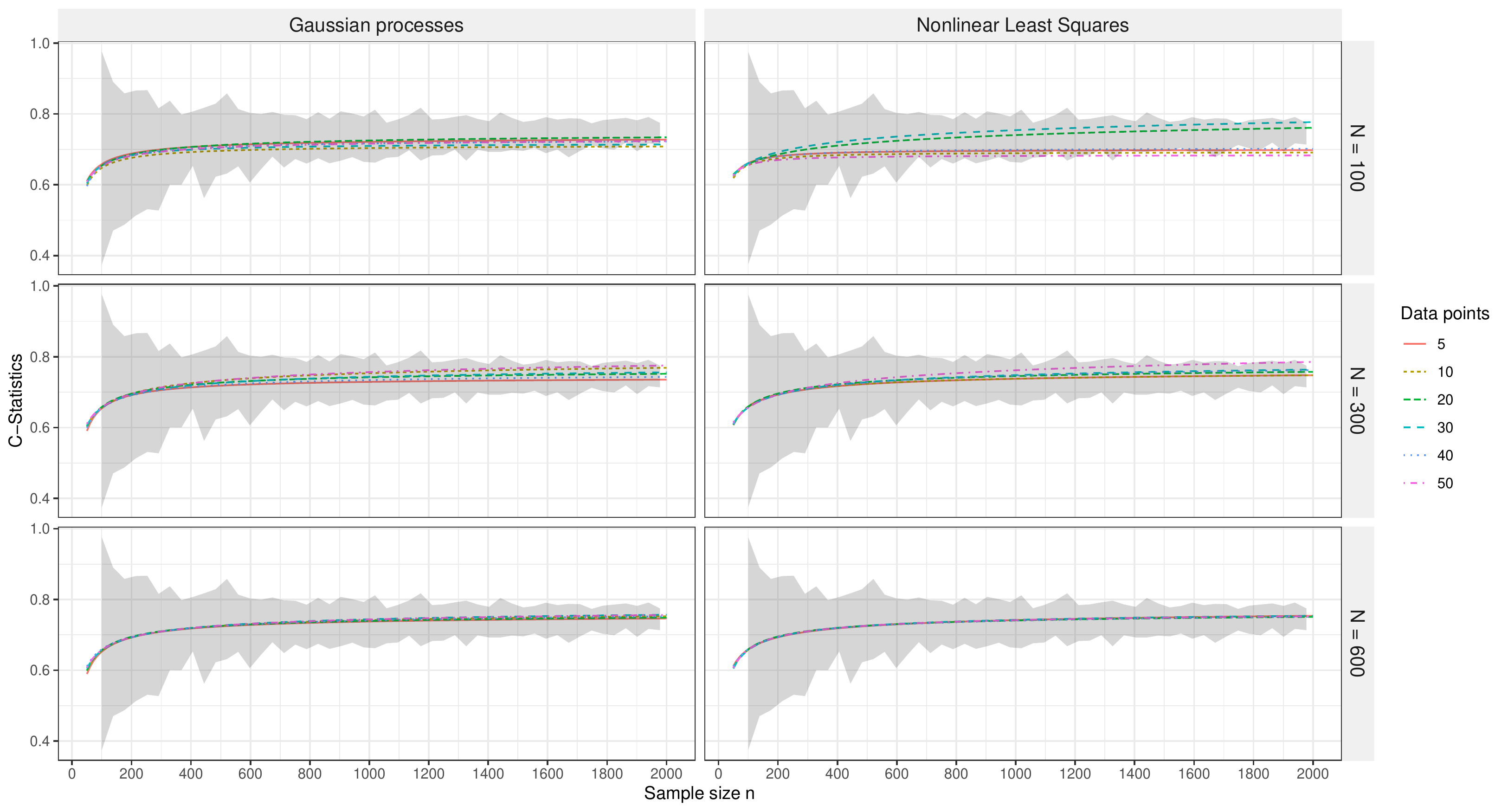}
  \captionof{figure}{Predicted C-statistics of learning curve of METABRIC with GP and NLS of logistic regression with different data points and at different total sample size (lines) and 95\% confidence interval of raw bootstrap results using all samples (shaded area)}
  \label{fig:figs1}
\end{center}

\begin{center}
  \captionof{table}{Point estimation and standard error (SE) of GP and NLS with different data points and at different total sample size}
  \label{tab:tabs1}
  \footnotesize
  \begin{tabular}{lcccccc}
    \tblhead{
    & \multicolumn{3}{c}{\textbf{Gaussian Process}} & \multicolumn{3}{c}{\textbf{Nonlinear Least Squares}}\\
    \textbf{m} & \textbf{a (SE)} & \textbf{b (SE)} & \textbf{c (SE)} & \textbf{a (SE)} & \textbf{b (SE)} & \textbf{c (SE)}
    }
    \multicolumn{6}{l}{\textbf{N = 100}}  \\
    5 & 0.26 (0.09) & 1.77 (1.73) & 0.67 (0.24) & 0.30 (0.76) & 3.88 (36.44) & 1.00 (594401.73)\\
    10 & 0.29 (0.02) & 2.68 (1.05) & 0.80 (0.11) & 0.31 (0.59) & 3.71 (28.94) & 1.00 (482237.84)\\
    20 & 0.25 (0.02) & 1.22 (0.24) & 0.54 (0.07) & 0.00 (145040511.44) & 0.59 (5.66) & 0.12 (22.14)\\
    30 & 0.28 (0.02) & 3.24 (0.92) & 0.85 (0.10) & 0.00 (2337166.71) & 0.65 (3.13) & 0.14 (14.03)\\
    40 & 0.27 (0.02) & 2.80 (1.23) & 0.78 (0.17) & 0.29 (0.52) & 1.94 (8.54) & 0.81 (9.56)\\
    50 & 0.26 (0.02) & 1.38 (0.24) & 0.61 (0.06) & 0.32 (0.30) & 2.89 (15.25) & 1.00 (170639.92)\\
    \multicolumn{6}{l}{\textbf{N = 300}}  \\
    5 & 0.26 (0.03) & 3.19 (1.53) & 0.77 (0.15) & 0.23 (0.25) & 1.34 (0.95) & 0.54 (0.97)\\
    10 & 0.18 (0.04) & 1.03 (0.31) & 0.40 (0.12) & 0.23 (0.19) & 1.37 (0.82) & 0.55 (0.80)\\
    20 & 0.23 (0.01) & 1.60 (0.30) & 0.57 (0.06) & 0.21 (0.20) & 1.22 (0.53) & 0.49 (0.61)\\
    30 & 0.21 (0.01) & 1.14 (0.13) & 0.47 (0.04) & 0.18 (0.38) & 0.85 (0.26) & 0.36 (0.61)\\
    40 & 0.24 (0.01) & 1.75 (0.41) & 0.61 (0.07) & 0.19 (0.29) & 0.87 (0.23) & 0.37 (0.51)\\
    50 & 0.14 (0.02) & 0.80 (0.04) & 0.30 (0.03) & 0.05 (3.40) & 0.71 (0.03) & 0.19 (0.69)\\
    \multicolumn{6}{l}{\textbf{N = 600}}  \\
    5 & 0.24 (0.03) & 2.29 (1.36) & 0.66 (0.16) & 0.21 (0.14) & 1.01 (0.33) & 0.45 (0.45)\\
    10 & 0.22 (0.02) & 1.50 (0.45) & 0.55 (0.09) & 0.22 (0.08) & 1.29 (0.37) & 0.51 (0.35)\\
    20 & 0.23 (0.01) & 1.81 (0.41) & 0.61 (0.07) & 0.23 (0.05) & 1.44 (0.35) & 0.55 (0.28)\\
    30 & 0.21 (0.01) & 1.13 (0.16) & 0.46 (0.05) & 0.22 (0.06) & 1.27 (0.26) & 0.51 (0.24)\\
    40 & 0.22 (0.00) & 1.11 (0.08) & 0.48 (0.02) & 0.22 (0.05) & 1.21 (0.23) & 0.51 (0.22)\\
    50 & 0.21 (0.01) & 1.08 (0.15) & 0.46 (0.05) & 0.22 (0.04) & 1.29 (0.22) & 0.52 (0.19)\\
    \lastline
    \end{tabular}
\end{center}

\end{document}